\documentclass[10pt, conference, letterpaper]{IEEEtran}
\IEEEoverridecommandlockouts
\usepackage[T1]{fontenc}
\usepackage[utf8]{inputenc}
\usepackage{amsfonts,amssymb}
\usepackage{algorithmic}
\usepackage{graphicx}
\usepackage{textcomp}
\usepackage{xcolor}
\usepackage{csquotes,rotating,tikz,enumitem,tabularx,xfrac,bm}
\usepackage{hyperref} 
\usepackage{multirow, makecell, siunitx, tabto}
\usepackage{subcaption}
\usepackage{uulm-logos}
\usepackage[linesnumbered,lined,boxed,commentsnumbered]{algorithm2e}
\usepackage[style=ieee,backend=biber,sortcites=true,minnames=5,maxnames=10,mincitenames=1,maxcitenames=2,bibencoding=utf8,
doi=false,isbn=false,natbib]{biblatex}
\AtEveryBibitem{%
	\clearlist{language}%
	\ifentrytype{misc}{%
	}{%
		\ifentrytype{techreport}{%
		}{%
			\clearfield{url}%
			\clearfield{urlyear}%
		}%
	}%
	\ifentrytype{inproceedings}{%
		\clearfield{pages}%
		\clearlist{location}%
		\clearfield{month}%
		\clearfield{series}%
	}{%
	}%
}
\def\BibTeX{{\rm B\kern-.05em{\sc i\kern-.025em b}\kern-.08em
		T\kern-.1667em\lower.7ex\hbox{E}\kern-.125emX}}

\usetikzlibrary{positioning, calc, fit, backgrounds}
\bibliography{bibliography}

\newcommand{\hex}[1]{\textbackslash x#1}
\newcommand{\webref}[1]{\footnote{\url{#1}}}

\usepackage[textsize=footnotesize, disable]{todonotes}  
\usepackage{xargs}
\newcommandx{\todorens}[2][1=]{\todo[linecolor=blue,backgroundcolor=blue!25,bordercolor=blue,#1]{#2}}

\usepackage{amsmath}
\DeclareMathOperator*{\argmax}{arg\,max}

\newcommand{\nemetyl}{NEMETYL\xspace}

\newcommand{\byte}{\ensuremath{b}} 
\newcommand{\seg}{\ensuremath{s}} 
\newcommand{\segalt}{\ensuremath{t}} 
\newcommand{\feature}{\ensuremath{\mathbf{s}}} 
\newcommand{\featurealt}{\ensuremath{\mathbf{t}}} 
\newcommand{\candist}{\ensuremath{d_C}} 
\newcommand{\mincandist}{\ensuremath{d_\beta}} 
\newcommand{\mixedcandist}{\ensuremath{d_m}} 
\newcommand{\len}[1]{\ensuremath{|#1|}} 
\newcommand{\msglen}[1]{\ensuremath{|#1|_s}} 

\newcommand{\canberrapenalty}{\ensuremath{p_f}} 

\newcommand{\gap}{\ensuremath{p_g}}
\newcommand{\match}{\ensuremath{p_m}}
\newcommand{\mismatch}{\ensuremath{p_d}}
\newcommand{\nwmin}{\ensuremath{\mathfrak{t}_{\min}}}
\newcommand{\nwmax}{\ensuremath{\mathfrak{t}_{\max}}}

\newcommand{\tmin}{\ensuremath{t_l}} 
\newcommand{\tmax}{\ensuremath{t_h}} 

\newcommand{\msg}{\ensuremath{m}} 
\newcommand{\msgset}{\ensuremath{M}} 

\newcommand{\nwdist}{\ensuremath{d_{\text{NW}}}} 
\newcommand{\nwscore}{\ensuremath{\mathfrak{N}}} 

\begin{document}
	
	\title{
		Message Type Identification of Binary Network Protocols using Continuous Segment Similarity\\
	}
	
	\author{\IEEEauthorblockN{Stephan Kleber}
		\IEEEauthorblockA{\textit{Institute of Distributed Systems} \\
			\textit{Ulm University}, Germany \\
			stephan.kleber@uni-ulm.de}
		\and
		\IEEEauthorblockN{Rens W. van der Heijden}
		\IEEEauthorblockA{\textit{Institute of Distributed Systems} \\
			\textit{Ulm University}, Germany \\
			rensvdheijden@gmail.com}
		\and
		\IEEEauthorblockN{Frank Kargl}
		\IEEEauthorblockA{\textit{Institute of Distributed Systems} \\
			\textit{Ulm University}, Germany \\
			frank.kargl@uni-ulm.de}
	}
	  
	\maketitle
	
    \IEEEpubid{\begin{minipage}{\textwidth}\ \\[24pt]
        \copyright 2020 IEEE. Personal use of this material is permitted. Permission from IEEE must be
        obtained for all other uses, in any current or future media, including
        reprinting/republishing this material for advertising or promotional purposes, creating new
        collective works, for resale or redistribution to servers or lists, or reuse of any copyrighted
        component of this work in other works.\\
        IEEE Conference on Computer Communications 2020.
        DOI: \href{https://doi.org/10.1109/INFOCOM41043.2020.9155275}{10.1109/INFOCOM41043.2020.9155275}
    \end{minipage}} 
    
    \IEEEpubidadjcol
	\begin{abstract}
		Protocol reverse engineering based on traffic traces infers the behavior of unknown network protocols by analyzing observable network messages.
        To perform correct deduction of message semantics or behavior analysis, accurate message type identification is an essential first step.
        However, identifying message types is particularly difficult for binary protocols,
		whose structural features are hidden in their densely packed data representation.
		We leverage the intrinsic structural features of binary protocols and propose an accurate method for discriminating message types.
		
		Our approach uses a similarity measure with continuous value range by comparing feature vectors where vector elements correspond to the fields in a message, rather than discrete byte values.
		This enables a better recognition of structural patterns, which remain hidden when only exact value matches are considered.
		We combine Hirschberg alignment with DBSCAN as cluster algorithm to yield a novel inference mechanism.
        By applying novel autoconfiguration schemes, we do not require manually configured parameters for the analysis of an unknown protocol, as required by earlier approaches.
		
		Results of our evaluations show that our approach has considerable advantages in message type identification result quality and also execution performance over previous approaches.
	\end{abstract}
	

\section{Introduction}\label{sec:introduction}


Several recent surveys by~\citet{narayan_survey_2015, duchene_state_2018, kleber_survey_2019} describe the current state of the art for protocol reverse engineering based on network traffic traces or programs.
In this paper, we focus on traffic analysis based on information gained by observing only the communication link.
Such traffic analysis is non-invasive and does not require access to programs or control over any entity and therefore is regularly applied~\cite{capec_content_team_capec_2014}.
This method of network analysis has been used to gain comprehension of hitherto unknown network protocol,
e.\,g., 
NetBios services~\cite{leita_scriptgen:_2005}, the Koobface command-and-control protocol~\cite{krueger_learning_2012}, OSCAR~\cite{gascon_pulsar:_2015}, and IoT protocols~\cite{wressnegger_zoe:_2018}.
The knowledge gained was then used for network analysis and security-relevant tasks, like
vulnerability testing by fuzzing~\cite{gascon_pulsar:_2015},
the setup of honeypots~\cite{leita_scriptgen:_2005, krueger_learning_2012},
analyzing botnets~\cite{krueger_learning_2012},
and automated network modeling~\cite{wressnegger_zoe:_2018}.

\begin{figure}
	\hspace{-1ex}
	\begin{tikzpicture}[
	every node/.style={outer sep=.5ex, font=\sffamily\footnotesize}
	]
	\coordinate (top) at (0, 1.7em);
	\coordinate (bot) at (0,-1.7em);
	
	\node[yshift=-.2ex] at (0,0) (ppr) {Preprocessing\phantom{g}};
	\node[right=1em of ppr] (str) {Structure\phantom{g}};
	\node[right=1em of str] (nemetyl) {\nemetyl\phantom{g}};
	\node[right=1em of nemetyl] (grm) {\phantom{j}Grammar};

	\begin{scope}[on background layer, 
	every path/.style={transform canvas={xshift={.5ex}}, fill=uulm}
	]
	\foreach \x in {ppr} {
		\path[fill=uulm-in!60]
		(\x.west |- top) to (\x.west |- bot)
		to ([xshift=-2em]\x.east |- bot) to (\x.east)
		to ([xshift=-2em]\x.east |- top) to cycle;
	}
	\foreach \x in {str} {
		\path[fill=uulm-in!60]
		([xshift=-2em]\x.west |- top) to (\x.west) to
		([xshift=-2em]\x.west |- bot)
		to ([xshift=-2em]\x.east |- bot) to (\x.east)
		to ([xshift=-2em]\x.east |- top) to cycle;
	}
	\foreach \x in {nemetyl} {
		\path
		([xshift=-2em]\x.west |- top) to (\x.west) to
		([xshift=-2em]\x.west |- bot)
		to ([xshift=-2em]\x.east |- bot) to (\x.east)
		to ([xshift=-2em]\x.east |- top) to cycle;
	}
	\foreach \x in {grm} {
		\path[fill=uulm-in!60]
		([xshift=-2em]\x.west |- top) to (\x.west)
		to ([xshift=-2em]\x.west |- bot)
		to (\x.east |- bot)
		to (\x.east |- top) to cycle;
	}
	\end{scope}

	\begin{scope}[
	every node/.style={align=left, anchor=west, font=\sffamily\footnotesize},
	goa/.style={->, line width=2pt, transform canvas={xshift=2ex}, uulm-akzent},
	rotup/.style={rotate=10, anchor=south west},
	rotdn/.style={rotate=-10, anchor=north west}
	]
	\coordinate (goa) at (0, -2.2em);
	\node[rotdn] (pprgoa) at (goa -| ppr.west) {Recorded\\message trace};
	\node[rotdn] (strgoa) at (goa -| str.west) {Message\\formats};
	\node[rotdn] (nemetylgoa) at (goa -| nemetyl.west) {Message\\types};
	\node[rotdn] (grmgoa) at (goa -| grm.west) {State machine};
	
	\foreach \x/\y in {ppr/pprgoa, str/strgoa, nemetyl/nemetylgoa, grm/grmgoa} {
		\draw[goa] (\x.south west) to (\y.north west);
	}
	\end{scope}
	\end{tikzpicture}
	\caption{Steps of network traffic trace analysis.}
	\label{fig:pre-context}
\end{figure}
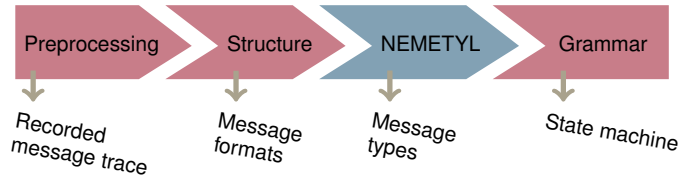
Protocol reverse engineering based on network traffic traces can roughly be divided into four steps, as shown in \autoref{fig:pre-context}.
First, the protocol traffic is recorded into traces during pre-processing.
The second step is a structural analysis, whose purpose is to derive formats (e.\,g., data fields) of individual messages exchanged within the protocol.
After analyzing the internal structure of messages, one important step is grouping these messages into message types (for DHCP, e.\,g., Discover, Offer, ACK). 
This information can then be used to derive a state machine that describes the grammar of the protocol.
We place the derivation of the state machine out of the scope of our work, since adequate methods already exist for this analysis step~\cite{antunes_reverse_2011, krueger_learning_2012, bossert_towards_2014, goo_protocol_2019} that can be combined with our work.
In this paper, we propose a novel approach to automate the task of message type identification for binary protocols.


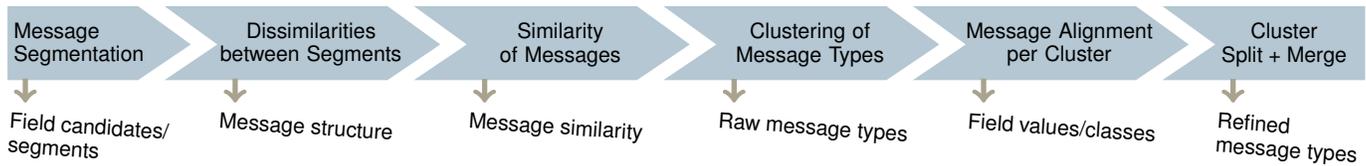
\begin{figure*}[t]
	\hspace{-1em}
	\begin{tikzpicture}[
	every node/.style={outer sep=.3ex, align=center, font=\sffamily\footnotesize},
	fixedwidth/.style={text width=7.5em, transform canvas={xshift=-1ex}}
	]
	\coordinate (top) at (0, 1.7em);
	\coordinate (bot) at (0,-1.7em);
	
	\node[align=left, text width=6em] (toc) at (0,0) {Message\\Segmentation\vphantom{g}};
	\node[right=1em of toc, fixedwidth] 
	(sim) {Dissimilarities between Segments\vphantom{g}};
	\node[right=1em of sim, fixedwidth] (sis) {Similarity\\ of Messages\vphantom{g}};
	\node[right=1em of sis, fixedwidth] (cst) {Clustering of\\ Message Types\vphantom{g}};
	\node[right=1em of cst, fixedwidth] (agn) {Message Alignment\\ per Cluster\vphantom{g}};
	\node[right=1em of agn] (fse) {Cluster\\ Split + Merge\vphantom{g}};
	
	\begin{scope}[on background layer, 
	every path/.style={transform canvas={xshift={.5ex}}, fill=uulm!60}
	]
	\foreach \x in {toc} {
		\path
		(\x.west |- top) to (\x.west |- bot)
		to ([xshift=-2em]\x.east |- bot) to (\x.east)
		to ([xshift=-2em]\x.east |- top) to cycle;
	}
	\foreach \x in {sim, sis, agn, cst} {
		\path
		([xshift=-2em]\x.west |- top) to (\x.west) to
		([xshift=-2em]\x.west |- bot)
		to ([xshift=-2em]\x.east |- bot) to (\x.east)
		to ([xshift=-2em]\x.east |- top) to cycle;
	}
	\foreach \x in {} {
		\path[fill=uulm-in]
		([xshift=-2em]\x.west |- top) to (\x.west) to
		([xshift=-2em]\x.west |- bot)
		to ([xshift=-2em]\x.east |- bot) to (\x.east)
		to ([xshift=-2em]\x.east |- top) to cycle;
	}
	\foreach \x in {fse} {
		\path
		([xshift=-2em]\x.west |- top) to (\x.west)
		to ([xshift=-2em]\x.west |- bot)
		to (\x.east |- bot)
		to (\x.east |- top) to cycle;
	}
	\end{scope}

	\begin{scope}[
	every node/.style={align=left, anchor=west, font=\sffamily\footnotesize},
	goa/.style={->, line width=2pt, transform canvas={xshift=2ex}, uulm-akzent},
	rotup/.style={rotate=4, anchor=south west},
	rotdn/.style={rotate=-4, anchor=north west}
	]
	\coordinate (goa) at (0, -2.2em);
	\node[rotdn] (tocgoa) at (goa -| toc.west) {Field candidates/\\segments};
	\node[rotdn] (simgoa) at (goa -| sim.west) {Message structure};
	\node[rotdn] (sisgoa) at (goa -| sis.west) {Message similarity};
	\node[rotdn] (agngoa) at (goa -| agn.west) {Field values/classes};
	\node[rotdn] (cstgoa) at (goa -| cst.west) {Raw message types};
	\node[rotdn] (fsegoa) at (goa -| fse.west) {Refined\\ message types};
	
	\foreach \x/\y in {toc/tocgoa, sim/simgoa, agn/agngoa, cst/cstgoa, fse/fsegoa, sis/sisgoa} {
		\draw[goa] (\x.south west) to (\y.north west);
	}
	\end{scope}
	
	\end{tikzpicture}
	\caption{Sequence of steps of \nemetyl and its intermediate results.}
	\label{fig:nemetyl-steps}
\end{figure*}


\IEEEpubidadjcol

As the mentioned surveys show, current methods for traffic analysis mainly focus on textual protocols, which use separators and keywords that are discernible by natural language processing (NLP).
Most binary network protocols, which represent data in concise bit patterns, lack these structural features required for NLP.
Most of the few known methods tailored for binary protocols
are derived from bio-informatics algorithms, e.\,g., Needleman-Wunsch~\cite{needleman_general_1970}.
When analyzing network protocols by these algorithms, messages commonly are interpreted as sequences of bytes.
%
To work around the exponential complexity of na\"ive multiple sequence alignment~\cite{wang_complexity_1994}, 
known applications for message analysis use the agglomerative clustering of UPGMA~\cite{sokal_statistical_1958}.
%
The algorithm recursively \textit{agglomerates}, i.\,e. merges, similar pairs of clusters.
This process builds a phylogenetic guide-tree with a complexity of the square of the amount of sequences~\cite{wang_complexity_1994}.

In the bio-informatics use case, a phylogenetic tree is favorable, since it reflects the evolution of genome sequences.
For protocol messages, no evolutionary relations exist, thus the computational overhead of agglomerative clustering is unnecessary.
Moreover, aligning individual byte values detaches them from their context.
This loss of context is undesirable since 
the exact same value cannot be unconditionally expected in multiple independent messages of binary protocols;
this may lead to spurious relationships of values across messages.

In contrast, our segment-based approach is designed to provide better efficiency while improving the result quality at the same time.
To accomplish this, instead of aligning byte-by-byte, we first split messages into segments using computationally cheap heuristics, like cutting them into fixed length chunks or applying a more advanced approach called NEMESYS~\cite{kleber_nemesys:_2018}.
Based on domain knowledge about the characteristics of typical network messages, we then derive message types from these segmented messages.
Since we deal with unknown protocols, structure and data values are obscured.
Thus, like previous approaches, we must rely on general domain knowledge about network protocols.
We obtained this knowledge from analyzing numerous binary protocols in order to ensure our approach's general applicability.
From this knowledge, we derive assumptions that allow us to design heuristic methods on which our approach relies.
These heuristics do not rely on phylogenetic relations, enabling us to use a more efficient clustering algorithm.

The main contribution of our paper is a novel \emph{message type identification} approach for unknown binary protocols.
Unlike previous methods, our approach is designed to provide higher accuracy without manual parameter selection.
To achieve this goal, we developed solutions to task-specific but fundamental challenges in the handling of sequential binary data.
These contributions are 
the interpretation of \emph{binary data as feature vectors},
a novel way to apply a \emph{vector distance to unequally-sized feature vectors},
the application of \emph{Hirschberg alignment} and the \emph{DBSCAN cluster algorithm} to the area of protocol reverse engineering,
and a novel method for \emph{auto-configuring the DBSCAN parameter $\epsilon$}.
We implement and evaluate our approach in \textbf{\nemetyl}:
\textbf{NE}twork \textbf{ME}ssage \textbf{TY}pe identification by a\textbf{L}ignment on similarity between message segment feature vectors.
To evaluate \nemetyl we use known real-world binary protocols as a baseline.
We determine the inference quality in terms of well-known cluster properties and compare them to the byte-wise sequence-alignment message inference performed by the state-of-the-art tool Netzob\footnote{\url{github.com/netzob/netzob}. All URLs last accessed on 18 Dec 2019.}~\cite{bossert_towards_2014}.

\medskip
In the next section, we discuss related work.
In \autoref{sec:approach}, we present the details of our approach followed by 
a description of design decisions for its implementation \nemetyl in \autoref{sec:implementation}.
\autoref{sec:evaluation} contains the results of our evaluation of \nemetyl.
Finally, we outline our ideas for future work and conclude the paper in Sections \ref{sec:futurework} and \ref{sec:conclusion}.

\section{Related Work}\label{sec:related-work}
Static traffic analysis is a specific kind of protocol reverse engineering, where network traffic between two genuine entities is monitored purely passively, e.g.,  without injecting own messages or filtering some. 
The use of sequence alignment to perform such an analysis was first suggested by \citet{beddoe_network_2004}.
Since \citeauthor{beddoe_network_2004}'s paper, a variety of algorithms from natural language processing and bio-informatics have been applied to network protocols~\cite{kleber_survey_2019}.
There are also practical implementations to perform static traffic analysis, the most versatile of which is Netzob~\cite{bossert_towards_2014}.
%



ScriptGen \cite{leita_scriptgen:_2005}, 
Discoverer \cite{cui_discoverer:_2007}, and 
Netzob all use sequence alignment to infer message types.  
ScriptGen and Discoverer differ from Netzob in that they align subsequences of messages (tokens or segments) instead of single bytes.
They propose effective methods to identify such tokens in textual protocols.
In addition, ScriptGen proposes the derivation of tokens from frequency, variance, and other byte characteristics throughout all messages of a trace.
Although ScriptGen and Discoverer claim their methods to be universally applicable, they leave it to future work to solve the details of analyzing binary protocols.
All presented methods are 
using agglomerative clustering by UPGMA~\cite{sokal_statistical_1958},
which prohibits the analysis of large traces and ignores byte-contexts as discussed in the introduction.

FieldHunter \cite{bermudez_towards_2016}
combines concepts from Netzob, Discoverer, ScriptGen, and other related methods.
It provides solutions for a number of challenges for format inference, like characterization of field types.
However, FieldHunter is still based on byte-value features and thus also misses details related to message structure.

In related work, the coverage of the analysis of binary protocols is sparse and the robustness of existing solutions is limited due to the kind of byte-wise analyses they perform.
Furthermore, the common combination of agglomerative clustering with conventional sequence alignment prohibits the analysis of large traces due to the high computational demand.



\section{Approach}\label{sec:approach}
Our approach, \nemetyl, consists of the steps illustrated in \autoref{fig:nemetyl-steps}.
These steps are described in the following subsections.
Among others, we require a suitable definition of similarity between segments (message subsequences) as well as between messages (message structure) and
then use clustering to identify messages with a similar structure.
Finally, we refine these clustered message types to take into account differences between types that have the same structure.

\subsection{Message Segmentation}
As a prerequisite, \nemetyl relies on a segmentation of each message into atomic chunks of consecutive bytes.
Ideally, such segments correspond to protocol fields, but for \nemetyl an approximate match of segments with field boundaries suffices.
Thus, any heuristic method to obtain segments from messages may be applicable.
In our current work, we investigate three different segmenters:
\begin{description}[noitemsep, nolistsep]
	\item[tshark] generates segments from true fields determined by tshark's dissectors.
		This qualifies as ground truth for field boundaries of messages from already known protocols.
	
	\item[4-bytes-fixed] uses fixed chunks of 4 byte length as segments.
		With no additional information available about a protocol to analyze, this is an extremely simple and efficient fallback that is always available.
		
	\item[NEMESYS] is a message segmenter 
		we proposed previously~\cite{kleber_nemesys:_2018}.
		This heuristic and computationally cheap method approximates field boundaries for any unknown network protocol by information-theoretical metrics.
\end{description}

\subsection{Dissimilarity between Segments}
To quantify dissimilarity between segments, we first need a representation that describes the contents of each segment.

\medskip
\subsubsection{Extracting Feature Vectors from Message Segment Bytes}
We generate feature vectors from each segment's byte values,
use hexadecimal notation for bytes, e.\,g., $\byte = \mathtt{0xA4}$, 
and write a segment $\seg$ as an ordered set of $n$ byte values $\byte_0, \dots, \byte_{n-1}$:
$\seg = \langle b_i \rangle \mathrm{, with\ } i \in \left[0,n-1\right]$


%
We interpret the sequence of byte values $\langle \byte_i \rangle$ of a segment as a feature vector $\feature$, where the $i$th vector component $\feature_i=\byte_i$.
For example, given a segment $\seg = \langle \byte_i \rangle = \langle \text{\texttt{0x17, 0x23, 0x00, 0x42}} \rangle$, the feature vector is written as $\feature = (\texttt{0x17}, \texttt{0x23}, \texttt{0x00},\texttt{0x42})^T$.

\medskip
\subsubsection{Canberra Dissimilarity}
We use the Canberra distance \cite{lance_computer_1966} to quantify the dissimilarity between segments, which for vectors $u$ and $v$ of equal dimension $n$, is defined as:
\begin{equation}
  \candist(u,v) = \sum^{n-1}_{i=0} \frac{|u_i-v_i|}{|u_i| + |v_i|},
\end{equation}
The Canberra distance correlates well with the intuitive similarity of byte sequences in that it relates different values in sequences to their common mean of values.
However, our application as a measure of dissimilarity requires to extend this distance to be applicable to vectors of differing dimensionality; 
otherwise segments of unequal length would not be comparable.
To accomplish this, we \emph{generalize the notion of the Canberra distance to vectors of different length} by choosing an embedding of the higher-dimensional vector into the lower-dimensional space. 
We normalize the distances from the embedded vectors to assure comparability across vector spaces.
Note that this generalization is only valid for our specific application in which the vector represents a segment, as it violates the triangle inequality, and therefore constitutes not a distance anymore, but instead a dissimilarity.
We call this the \emph{Canberra dissimilarity} $d_m$.
%

Without loss of generality, let $\seg$ and $\segalt$ be segments with $\len{\seg}\le\len{\segalt}$ and generate the associated vectors $\seg\rightarrow\feature$, $\segalt\rightarrow\featurealt$.
The lengths of the segments imply the dimensionality of the associated vectors: $\len{\seg} = \dim{\feature}$ and $\len{\segalt} = \dim{\featurealt}$.
For the case where $\len{\seg}=\len{\segalt}$, we define $\mixedcandist(\feature,\featurealt)=\frac{\candist(\feature,\featurealt)}{\len{\seg}}$.
For the case where $\len{\seg}<\len{\segalt}$, we apply the following procedure to derive the Canberra dissimilarity.

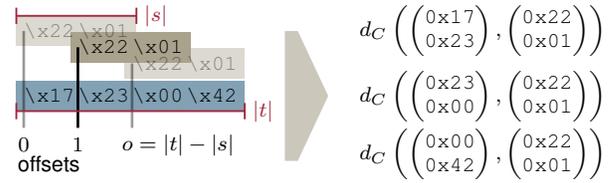
\begin{figure}
	\centering
	\footnotesize
	\begin{tikzpicture}[
		xscale=.2ex, yscale=.06ex,
		hex/.style={font=\ttfamily},
		iter/.style={opacity=.4},
		len/.style={uulm-in, line width=.7pt},
		axshift/.style={xshift=3pt},
		ticks/.style={line width=1pt, transform canvas={xshift=3pt}},
	]
	\node[hex] (Oa) at (0,0) {\hex{17}};
	\node[hex] (Ob) at (1,0) {\hex{23}};
	\node[hex] (Oc) at (2,0) {\hex{00}};
	\node[hex] (Od) at (3,0) {\hex{42}};
	
	\node[hex, iter] (Ia) at (0,4) {\hex{22}};
	\node[hex, iter] (Ib) at (1,4) {\hex{01}};
	
	\node[hex] (Ic) at (1,3) {\hex{22}};
	\node[hex] (Id) at (2,3) {\hex{01}};
	
	\node[hex, iter] (Ie) at (2,2) {\hex{22}};
	\node[hex, iter] (If) at (3,2) {\hex{01}};
	
	\draw[len] (Ia.north west) to (Ib.north east);
	\draw[len] (Ia.north west)+(0,-.5) to +(0,.5);
	\draw[len] (Ib.north east)+(0,-.5) to +(0,.5);
	\node[anchor=west, len] at (Ib.north east) {$ \left| \seg \right| $};
	
	\draw[len] (Oa.south west) to (Od.south east);
	\draw[len] (Oa.south west)+(0,-.5) to +(0,.5);
	\draw[len] (Od.south east)+(0,-.5) to +(0,.5);
	\node[anchor=west, len] at (Od.south east) {$ \left| \segalt \right| $};
	
	\begin{scope}[on background layer, inner sep=0]
		\node[fit=(Oa)(Od), fill=uulm] (Ofill) {};
		\node[fit=(Ia)(Ib), fill=uulm-akzent, opacity=.4] (Iafill) {};
		\node[fit=(Ic)(Id), fill=uulm-akzent] (Icfill) {};
		\node[fit=(Ie)(If), fill=uulm-akzent, opacity=.4] (Iefill) {};
		
		\coordinate (offset) at ($(Ofill.south)-(0,1)$);
		
		\draw[ticks, iter] (Iafill.west) to (Iafill.west|-offset);
		\draw[ticks] (Icfill.west) to (Icfill.west|-offset);
		\draw[ticks, iter] (Iefill.west) to (Iefill.west|-offset);
	\end{scope}
	
	\coordinate (olabel) at (0,-3);
	\node[axshift] (label0) at (Iafill.west|-olabel) {$0$};
	\node[axshift] at (Icfill.west|-olabel) {$1$};
	\node[anchor=west, xshift=-1ex] at (Iefill.west|-olabel) 
		{$ o = \left| \segalt \right| - \left| \seg \right| $};
	\node[font=\sffamily, yshift=-2ex, anchor=west, xshift=-1.4ex] at (label0) {offsets};
	
	\coordinate (top) at (0,4);
	\coordinate (bot) at (0,-4);
	\node[minimum width=2em] (cor) at (4.75,0) {};
	
	\fill[uulm-akzent!60] 
		(cor.west |- top) to (cor.west |- bot)
		to ([xshift=-2em]cor.east |- bot) to (cor.east)
		to ([xshift=-2em]cor.east |- top) to cycle;
	
	\begin{scope}[yscale=.4ex, shift={(8,0)}]
	\node (v0) at (0,2.8) {$
	d_C\left(
	\begin{pmatrix}
	\texttt{0x17}\\
	\texttt{0x23}\\
	\end{pmatrix},
	\begin{pmatrix}
	\texttt{0x22}\\
	\texttt{0x01}\\
	\end{pmatrix}
	\right)
	$};

	\node (v1) at (0,0) {$
	d_C\left(
	\begin{pmatrix}
	\texttt{0x23}\\
	\texttt{0x00}\\
	\end{pmatrix},
	\begin{pmatrix}
	\texttt{0x22}\\
	\texttt{0x01}\\
	\end{pmatrix}
	\right)
	$};

	\node (v2) at (0,-2.5) {$
	d_C\left(
	\begin{pmatrix}
	\texttt{0x00}\\
	\texttt{0x42}\\
	\end{pmatrix},
	\begin{pmatrix}
	\texttt{0x22}\\
	\texttt{0x01}\\
	\end{pmatrix}
	\right)
	$};

	\end{scope}
	
%
%
	
	\end{tikzpicture}
	\caption{Sliding window interpretation of linear subspaces.}
	\label{fig:slide-vector-window}
\end{figure}

First, using a sliding window approach, we derive a series of possible embeddings of the longer feature vector $\featurealt$ into the vector space of the shorter feature vector.
The intuition behind this is illustrated in \autoref{fig:slide-vector-window}.
This shortened vector $\mathbf{\featurealt'}$ can then be compared using the standard Canberra distance, $\candist(\mathbf{\featurealt'}, \feature)$.
The sliding window of length $\len{\seg}=\dim{\feature}$ can be represented using an offset $o$ as $\featurealt_{[o,o+\len{\seg}]}$.
Using this notation, the set of suitable embeddings is defined as:
\begin{equation}
  T = \cup^{\len{\segalt}-\len{\seg}}_{o=0} \{ \featurealt_{[o,o+\len{\seg}]}\}
\end{equation}

Second, we define the minimum Canberra dissimilarity $\mincandist$ between a set of vectors $T$ as defined above and a shorter candidate vector $\feature$ as follows:
\begin{equation}
  \mincandist(T,\feature) = \frac{\min_{T}(\{\candist(\featurealt_{[o,o+\len{\seg}]})\})}{\len{\seg}}
\end{equation}
This selects the lowest Canberra distance as a dissimilarity between the longer and the shorter segment.
Essentially, we derive the nearest candidate vector (in terms of $\candist$).
We then normalize $\mincandist$ to $\len{\seg}$ in accordance with the explanation above.

Third, we note that for use as a dissimilarity, the pure $\mincandist$ provides undesirable results, since the difference in segment length is completely ignored.
Without the loss of generality, we therefore extend our previously defined dissimilarity $\mixedcandist$ with a non-linear modification based on the relative length of the segments.
Thus, we redefine%
\footnote{Our previous definition of $\mixedcandist$ still is valid as defined before, since $\mincandist(\feature, \featurealt) = \frac{\candist(\feature, \featurealt)}{\len{\seg}}$ and $r = 0$ for $\len{\seg}=\len{\segalt}$.}
$\mixedcandist$ as follows:
\begin{equation}
\mixedcandist(\feature, \featurealt) = 
\underbrace{
  \frac{\len{\seg}}{\len{\segalt}} \mincandist(\feature,\featurealt)
}_\text{subterm 1} +\\
\underbrace{r}_\text{subterm 2} + 
\underbrace{
  (1-\mincandist(\feature,\featurealt)) r (\frac{\len{\seg}}{\len{\segalt}^2} - \canberrapenalty)
}_\text{subterm 3}
\end{equation}
with $r=\frac{\len{\segalt}-\len{\seg}}{\len{\segalt}}$ representing the length difference between the segments, and with a parameter $\canberrapenalty$ to set the non-linear penalty in subterm 3, as discussed below.

\begin{table}[b]
	\centering
	\caption{Examples demonstrating the differences between minimum Canberra dissimilarity and Canberra dissimilarity.}\label{canberra-examples}
	\begin{tabular}{l|l|l|l|l}   
		Example & 1 & 2&3&4\\\hline
		$\featurealt$                   & \texttt{0x0008}  & \texttt{0x0208}            & \texttt{0x5706906e} & \texttt{0x0208} \\
		$\featurealt_{[o,\len{\seg}]}$  & \texttt{0x00}    & \texttt{\hphantom{00}0x08} & \texttt{0x5706}     & \texttt{0x0208} \\
		$\feature$                      & \texttt{0x00}    & \texttt{\hphantom{00}0x07} & \texttt{0x2700}     & \texttt{0x0008} \\
		$\mincandist$                   & 0.000            & 0.067                      & 0.690               & $d_m$ ($ \left| \seg \right| = \left| \segalt \right| $) \\ 
		$\mixedcandist$                 & 0.460            & 0.496                      & 0.814               & 0.5 \\
	\end{tabular}
\end{table}

The first subterm normalizes the minimum Canberra dissimilarity $\mincandist$ such that the longer segment's length $\len{\segalt}$ becomes the reference.
This represents the immediate dissimilarity between the selected components of the longer vector and all components of the shorter one.
As the dimensionality difference increases, more information is discarded from $\featurealt$ in the computation of $\mincandist$.
In \autoref{canberra-examples}, example 1 illustrates that $\mincandist$ erroneously assigns a distance of zero if the shorter vector is contained in the larger vector.
To correct for this, we use the second subterm, a linear factor $r$.
Example 2 in \autoref{canberra-examples} illustrates that $\mixedcandist$ attests two vectors with a small $\mincandist$ to have a dissimilarity just below their relative dimensionality difference ($\frac{1}{2}$).
Example 3 shows how this influences longer segment pairs that have the same relative dimensionality differences.
%
Finally, if a longer absolute part of $\segalt$ is used to calculate $\mixedcandist$, this should result in a lower dissimilarity of the segments, since we need to make less assumptions about the missing vector components.
For example, if 4 out of 5 components are selected by $\mincandist$, the final dissimilarity $\mixedcandist$ should be higher than if 16 of 20 components are known, even though the fraction is the same.
Intuitively, it is more unlikely that 16 components mismatch for any two similar segments than 4.
This absolute difference between the vectors' dimensionalities is taken into account by the third subterm, which non-linearly increases to penalize larger absolute dimensionality differences.
The factor $\canberrapenalty$ parameterizes the steepness of the slope in the penalty increase relative to $\mincandist$.

Each of the subterms describes a disjoined aspect of dissimilarity between two segments and we 
can, thus, connect them by simple summation.
The subterms representing these aspects sum up to a value of 1 for two maximally different segments.
$d_m$ decreases for more similar segments and reaches 0 for identical ones.


\medskip
\subsubsection{Refining Dissimilarities of Char Sequences}
Segments consisting of sequences of characters are special in a way that make the raw Canberra dissimilarity values between vectors for char sequences unsuitable.
We therefore reduce the dissimilarity values of pairs of segments that both are chars by a factor of 0.5 to reflect the characteristics of char sequences.
We explain our heuristic to identify char sequences in the implementation description (\autoref{sec:impl-segment-dissimilarities}).


\subsection{Similarity of Messages}\label{sub:msgsim}


In this step, we analyze the structural similarity of messages using the Needleman-Wunsch (NW) algorithm, which is widely used for this purpose in protocol reverse engineering.
The algorithm finds the optimal alignment of two strings using a substitution matrix $\mathbf{N}$ of possible alignments and the associated alignment score.
The maximum of these scores is the most suitable alignment.
The score is computed based on a gap penalty $\gap$ and a similarity matrix $\mathbf{S}$ that expresses the similarity of segments.
The gap penalty is used to penalize gaps (i.\,e., empty segments added to one of the messages) in the alignment.
The similarity between segments expresses whether they accurately align.
Most authors~\cite{beddoe_network_2004, leita_scriptgen:_2005, cui_discoverer:_2007, bossert_towards_2014} simplify the algorithm by choosing a similarity matrix such that the diagonal contains a match value $\match$ and all other values are $\mismatch$. 
This makes alignment a boolean decision, which is a major drawback of existing alignment methods~\cite{beddoe_network_2004}, as similar but unequal positions cannot be quantified appropriately.
In this section, we describe how the segment similarity of the previous section can be used to improve the alignment of NW.
We then use the NW score to define message similarity, from which we derive a message \textbf{dis}similarity $\nwdist$ that we use in the next step of \nemetyl.

%
%
%
%
%

\medskip
As an example, we use the two messages $\msg_0$ = \texttt{0x0208000807} and $\msg_1$ = \texttt{0x07270000082317}.
First, we need to derive the similarity matrix $S$ required for NW.
This is done by computing the pairwise similarity as $1-\mixedcandist$, where $\mixedcandist$ is the dissimilarity measure from the previous step.
Without loss of generality, we define that $\mismatch$ is the minimum similarity (zero) and $\match$ is the maximum similarity (one).
Intuitively, the pairwise similarity represents a spectrum of similarity, rather than the binary decision made by previous work.
For this example, this results in the following symmetric matrix:
\begin{center}
	\scriptsize
	\begin{tabular}{r|rrrrr}
		\textbf{$\mathbf{S}$} & \texttt{07} & \texttt{2700} & \texttt{2317} & \texttt{0208} & \texttt{0008} \\ \hline
		\texttt{07}   & 1.00 & 0.16 & 0.25 & 0.50 & 0.50 \\ 
		\texttt{2700} & 0.16 & 1.00 & 0.47 & 0.05 & 0.00 \\ 
		\texttt{2317} & 0.25 & 0.47 & 1.00 & 0.31 & 0.26 \\ 
		\texttt{0208} & 0.50 & 0.05 & 0.31 & 1.00 & 0.50 \\ 
		\texttt{0008} & 0.50 & 0.00 & 0.26 & 0.50 & 1.00 \\ 
	\end{tabular}
\end{center}

We use this matrix together with a gap penalty $\gap=-1$ to align the messages using NW.
This algorithm computes a substitution matrix $\mathbf{N}$ based on $\mathbf{S}$ and $\gap$, which for this example is:
\begin{center}
	\scriptsize
	\begin{tabular}{r|SSSS}
		$\mathbf{N}$  & & \multicolumn{1}{r}{\texttt{0208}} & \multicolumn{1}{r}{\texttt{0008}} & \multicolumn{1}{r}{\texttt{07}} \\ \hline
					  & 0 & -1 & -2 & -3 \\ 
		\texttt{07}   & -1 &  0.50 & -0.50 & -1.00 \\ 
		\texttt{2700} & -2 & -0.50 &  0.50 & -0.33 \\ 
		\texttt{0008} & -3 & -1.50 & -0.50 &  1.01 \\ 
        \texttt{2317} & -4 & -2.50 & -0.50 & \hfill \textbf{0.76}\\ 
	\end{tabular}
\end{center}

The NW score $\mathbf{N}_{[i,j]}$ in the substitution matrix is the degree of congruence of a specific pair of segments, namely the $i-1$th segment of $\msg_0$ (columns) and the $j-1$th of $\msg_1$ (rows).
Since the NW algorithm starts with blank strings, the dimensions of the matrix $\mathbf{N}$ are $\msglen{\msg_0}+1$ columns and $\msglen{\msg_1}+1$ rows, where $\msglen{\cdot}$ refers to the number of \emph{segments} 
with the matrix' indexing starting from 0. 
This matrix allows us to define the message similarity based on the NW score.
We follow \citet{smith_comparative_1981}, and define the NW score $\nwscore(\msg_0,\msg_1)$ as the bottom-right entry in the substitution matrix $\mathbf{N}$, i.e., $\nwscore(\msg_0,\msg_1)=\mathbf{N_{\msglen{\msg_0},\msglen{\msg_1}}}$.
The self-similarity of each message is the amount of segments, since in that case, each segment has a similarity of 1.
The message similarity can thus be described as follows:
\begin{table}[htbp]
	\centering
	\begin{tabular}{r|SS}
		 & \multicolumn{1}{c}{$\msg_0$} & \multicolumn{1}{c}{$\msg_1$}  \\ \hline
		$\msg_0$ & 4 & 0.76 \\ 
		$\msg_1$ & 0.76 & 3 \\ 
	\end{tabular}
\end{table}

NW scores quantify the similarity of messages through a value in $\mathbb{R}$.
However, our goal is to find groups of messages that are similar, for which an everywhere-positive dissimilarity function is required.
We thus define a function $\nwdist: \msgset \times \msgset \rightarrow \mathbb{R}^{+}$ based on the message similarity derived from the NW score $\nwscore(\msg_0,\msg_1)$:
\begin{equation}
  \nwdist(\msg_0, \msg_1) = 1 - \frac{\nwscore(\msg_0,\msg_1) - \nwmin(\msg_0,\msg_1)}{\nwmax(\msg_0,\msg_1) - \nwmin(\msg_0,\msg_1)}
\end{equation}
with $\nwmin (\msg_0,\msg_1)= \min(\msglen{\msg_0}, \msglen{\msg_1}) \cdot \min(\gap, \match, \mismatch)$ and $\nwmax(\msg_0,\msg_1) = \min(\msglen{\msg_0}, \msglen{\msg_1}) \cdot \max(\gap, \match, \mismatch)$.
Recall $\match$ and $\mismatch$ are bounds on segment similarity.
Therefore, this equation essentially normalizes the NW scores to a similarity value, which is then subtracted from 1 to gain a dissimilarity.

\medskip
By the end of this step, we have generated a message dissimilarity matrix that contains an entry for the dissimilarity between each pair of messages.

\subsection{Clustering of Message Types}
The message dissimilarity matrix from the previous step can directly be used as input to cluster messages into types.
The challenge is to automatically configure the parameters of the cluster method in a suitable way.
Our approach aims for a method that requires no a-priori knowledge of the protocol.
We identified DBSCAN \cite{ester_density-based_1996} to be suitable for this task.
It is a density-based method and therefore makes no assumptions about the shape of clusters.
DBSCAN requires no target number of clusters as parameter 
and deals with noise by rather not assigning a sample to any cluster than choosing the wrong one.
Applying clustering without prior knowledge about the trace requires to automatically derive the necessary parameters for DBSCAN from the trace itself.

\medskip
DBSCAN has two parameters, min\_samples and $\epsilon$, which are auto-configurable since they can be derived from statistical properties of the trace with a high rate of success.
The first, min\_samples, determines the smallest valid cluster 
(see \autoref{sec:impl-clustering}).
The second parameter, $\epsilon$, defines a range around a density core of samples that should constitute a cluster.
This parameter is specific to the data being clustered; in our case, it depends on the characteristics of a trace.
If $\epsilon$ is too large, random noise will be placed in clusters, while if $\epsilon$ is too small, meaningful clusters with only few samples will be considered noise.
As suitable values differ per trace, we developed an auto-configuration technique for this parameter.
%
Our choice for $\epsilon$ is based on the distance to the $k$th nearest neighbor, measured by NW-score dissimilarity $\nwdist$ between messages.
The intuition is that as this function typically shows a sudden change for a message set with well-defined clusters, the choice for $\epsilon$ should be exactly at this point.
The best choice for $\epsilon$ for the set of messages $\msgset$ in a trace thus depends on $k$, which is configured in the following.

We first define a family of functions $f_k: \msgset \rightarrow \mathbb{R}$ of which the $k$th function maps each message to the NW-score dissimilarity of the $k$th nearest neighbor of that message.
We refer to this as the $k$th nearest neighbor distance function, which quantifies how sparse a potential cluster of size $k$ is.
More formally,
\begin{align}
  f_k(\msg) = \max(D \subset \{ &\nwdist(\msg, \msg_k) : \msg_k \in \msgset \},\\\nonumber
  &|D| = k,  \text{s.t.} \sum_{a\in D} a \text{ is minimal})
\end{align}
We then apply a Gaussian filter (with parameter $\sigma$, see \ref{sec:impl-clustering}) and determine the inputs $(k, \msg)$ such that the curvature of the $k$ nearest neighbor distance function is maximal.
This process is biased towards large $k$, which we compensate for by dividing through the value of the function,
more formally:
\begin{equation}
  \argmax_{(k,\msg)} \frac{G(f_k''(m))}{G(f_k(m))}
\end{equation}
As $f_k$ is a family of discrete functions, \emph{curvature} is here defined using the forward difference, i.\,e., $f''_k(x) = f'_k(x+1) - f'_k(x)$ and $f'_k(x)=f_k(x+1)-f_k(x)$.
In other words, we find the $\msg_\epsilon$ and function $f_{k_\epsilon}$ for which the largest change in curvature is observed.
We then set $\epsilon$ to the value of the chosen function at the chosen message, i.\,e., $\epsilon=f_{k_\epsilon}(\msg_\epsilon)$.

\medskip
Finally, we cluster messages into types by DBSCAN using the auto-configured epsilon and the message dissimilarity matrix of all $\nwdist(\msg_i,\msg_j)$ as input.
%
This step yields highly precise clusters of messages classified into message types.

\subsection{Message Alignment per Cluster}
To determine the common internal structure for each message type, we align the messages within each cluster to the cluster's medoid, i.\,e.,
the message with the lowest dissimilarities to all other messages within the cluster.

This step results in clusters that each are commonly aligned on the segments of their messages.
Each common position in the alignment of all segments we call a \enquote{field candidate}.

\subsection{Cluster Refinement}
The steps up to this point -- segmentation, calculation of segment dissimilarity and message similarity, clustering, and alignment -- establish the structural similarities of messages and classify them accordingly.
Different message types that are structurally identical but distinguished only by dedicated field values, e.\,g., message type A is denoted by value \texttt{0x01} and type B by value \texttt{0x02} in field 1, cannot be discriminated by this process alone.
Therefore, we refine the clustering output by applying additional heuristics to find such cases, splitting underspecific clusters and merging overspecific clusters, to find the message types more accurately.
We discuss the details of these heuristics in \autoref{sec:impl-cluster-refinement}.

\medskip
\subsubsection{Splitting Underspecific Clusters}
To identify dedicated field values that distinguish structurally identical message types, we search for fields in the aligned clusters that contain only few distinct values.
These fields are likely to denote individual message types within one cluster and we therefore split it into sub-clusters that each contain only messages having one of the distinct values in this specific field.

\medskip
\subsubsection{Merging Overspecific Clusters}
For some structurally complex protocol messages, multiple clusters exist for a single message type.
We merge these overspecific clusters of similarly structured messages into one.
Thus, the alignment of each cluster is generalized into dynamic, static value (e.\,g., \texttt{0100}), and \texttt{GAP} field candidates.
Two clusters that exhibit a compatible structure according to the generalized alignment of the clusters are merged.
An example of the aligned field candidates for a mergeable pair of clusters looks like:
\begin{table}[h]
\centering
\begin{tabular}{cccccc}
\texttt{DYNAMIC} & \texttt{0100} & \texttt{0001} & \texttt{DYNAMIC} & \texttt{DYNAMIC} & \texttt{0001}\\
\texttt{DYNAMIC} & \texttt{0100} & \texttt{GAP } & \texttt{00     } & \texttt{001c   } & \texttt{0001}\\
\end{tabular}
\end{table}


\smallskip
The final result of our approach are clusters of messages with similar structure that have an increased accuracy compared to the result of the raw clustering.

\bigskip
\section{\nemetyl Implementation}\label{sec:implementation}
We implemented a modular proof of concept of \nemetyl to validate our approach.
%
It uses 
\todo{R2: used functionality}
numpy\footnote{\url{www.numpy.org}, version 1.13.3}, 
scipy\footnote{\url{www.scipy.org}, version 1.0.0}, sklearn.cluster.DBSCAN\footnote{\url{scikit-learn.org/0.19/modules/generated/sklearn.cluster.DBSCAN.html}}, 
netzob\footnote{\url{github.com/netzob/netzob/tree/next}, \enquote{next} branch}, 
and NEMESYS\webref{github.com/vs-uulm/nemesys}.
This section discusses parameter choices and design decisions made as part of the implementation and also highlights some more details on the general process.


%
%
%


\subsection{Message Segmentation}
We investigate three different segmentation approaches: \emph{tshark}, \emph{4-bytes-fixed}, and \emph{NEMESYS}.
The \emph{tshark} approach uses the true field boundaries and is thus not available for unknown protocols
but serves as a baseline for perfect segment knowledge. 
\emph{4-bytes-fixed} ignores real field boundaries and uses 4-byte chunks as segments.

Finally, \emph{NEMESYS} approximates the true field boundaries without knowledge of the true structure.
NEMESYS compares subsequent bytes with a method called \enquote{Gaussian-filtered Bit Congruence deltas}.
Throughout an individual message, this method reveals value patterns that it exploits to derive probable field boundaries.
In order to apply NEMESYS in \nemetyl, we needed to make three small modifications to the original scheme.
First, we additionally perform a frequency analysis of the most common segment values.
If one of these most frequent values is a subsequence of another, larger segment, we split this larger segment to reveal the more frequent and therefore more probable segment boundary.
Second, we enhance NEMESYS by our char sequence detection to improve the recognition of sequences of textual parts within the binary protocol.
Third, NEMESYS has issues with the initial segment of each message.
Therefore, we split each message's first segment into dedicated, one-byte segments.
Due to its heuristic nature, NEMESYS leads to less distinct message-type cluster borders than the tshark segmenter.
To compensate, we reduce the allowed range around density-cores for the DBSCAN-based clustering relative to the automatically determined value:
Therefore when using NEMESYS, we adjust $\epsilon$ by an additional factor of 0.8.

\subsection{Dissimilarities between Segments}\label{sec:impl-segment-dissimilarities}

As explained, we need to treat char sequences in a special way.
However, since our approach is intended for the analysis of unknown protocols, we have no definitive knowledge about where to expect char sequences.
We empirically determined characteristics of char sequences in 40\,000 messages of the protocols DNS, NBNS, SMB, and DHCP from real-world traces.
Thereby, we selected optimal values for a number of parameters that describe these characteristics to achieve a low number of false positives in the char detection.
Based on this analysis, we detect hypothetical char segments if all of the following conditions are true:
\begin{itemize}[noitemsep]  
	\item 
	All bytes have values lower than \texttt{0x7f}; 
\textit{(and)}
	\item 
	the sequence is at least 6 bytes long; 
\textit{(and)}
	\item 
	the mean of byte values is between the thresholds 
	$\tmin$ and 
	$\tmax$.
	We exclude zero bytes from the mean calculation to account for termination, padding,
	and latin chars in UTF-16.
	Our analysis determined working values to be $(\tmin, \tmax) = (50, 115)$;
\textit{(and)}
	\item 
	the ratio of non-printable and non-zero chars in one segment to the segment's length is below 0.33.
\end{itemize}
\medskip
\noindent
We further validated the char detection by applying this heuristic to NTP, which contains no chars, showing no false positives.

\subsection{Similarity of Messages}\label{sec:impl-message-similarities}
To compute NW scores, we use the Hirschberg alignment of the segments in a message pair.
Hirschberg alignment \cite{hirschberg_linear_1975} is a memory optimized version of Needleman-Wunsch \cite{needleman_general_1970}.
Our approach requires that a gap is only inserted in the alignment if no similar segments are available at one position.
Gaps therefore are scored negative and the value for the alignment parameter $p_g = -1$ was iteratively determined in a pilot study.



\subsection{Clustering of Message Types}\label{sec:impl-clustering}
To robustly handle noisy data for the auto-configuration of DBSCAN's $\epsilon$ parameter, we smooth the values of the discrete message distance function by a Gaussian filter $G(d(s_0, s_k))$.
The filter's single parameter is the standard deviation $\sigma$.
A pilot study showed that $\sigma = \ln(n)$ with $n$ being the number of neighbors removes noise sufficiently while retaining enough detail in the distance function.
%
We further limit the impact of noise in the dissimilarity values when selecting the distance function $f_k(m)$ by two actions:
First, we limit the iteration of $k$ to the closest 10\,\% of $k$-nearest neighbors.
Secondly, to remove boundary effects for values of $m$ close to the edges of the discrete distance function $f_k(m)$, we limit the range of $m$ to $2\sigma < m < n - 2\sigma$ after applying the Gaussian filter.


Since we merge overspecific clusters in the final step, we set DBSCAN's min\_samples parameter to the low value of $3$.

\subsection{Message Alignment per Cluster}
For the actual alignment of messages within a cluster, we reuse the fixed alignment parameters from \autoref{sec:impl-message-similarities}.
%
First, we ascendingly sort the messages in a cluster by their dissimilarity to the medoid.
%
Then we iterate through this list and align the next message to the common alignment of the medoid and all messages aligned in the previous iteration.
This constitutes a progressive alignment of the previous messages with the next one.
In each iterative alignment we introduce new gaps in all already aligned messages including the medoid.

\subsection{Cluster Refinement}\label{sec:impl-cluster-refinement}

\subsubsection{Splitting Underspecific Clusters}
To discriminate multiple message types within one cluster of structurally similar messages, we first need to determine which fields distinguish the message type.
As criterion, we use the observation that distinct values are frequent within one field that distinguishes the message type.
As a heuristic, $t = \lfloor \ln( \left| c \right| ) \rfloor $ is a threshold for the cluster $c$ to define a frequent value among $\left| c \right|$ values.
A field then is distinguishing message types if it contains only frequent values, determined by satisfying $\min(a) > t \geq \left| a \right| $ with $a$ being the sorted set of value counts of the field.
For example, if the values of one field in cluster $c$ are 5 times \texttt{85}, \texttt{01} is contained 8 times, and \texttt{23} 9 times,
then $a = \lbrace 5, 8, 9 \rbrace $, $\min(a) = 5$, $\left| a \right| = 3$, and $\left| c \right| = \sum a = 22$.
Therefore, $t$ is 3, and the inequality holds so that we consider this field to contain only frequent values.
Thus, we found a field that likely distinguishes message types.
%
All clusters will be split so that fields with frequent values will be the only value in the distinguishing field in a single cluster.

\medskip
\subsubsection{Merging Overspecific Clusters}
To determine if alignments are similar enough to merge clusters, we abstract the alignments: 
One sequence of abstracted field candidates represents the collective structure of the messages in a cluster.
We align these abstracted message structures of each pair of clusters by NW.
%
Two clusters are merged if at least one of the following is true for all aligned field candidates:

\medskip
\begin{itemize}[noitemsep]  
	\item A gap is present in either structure at one position.
	\item Both field candidates have the same static byte values or both are dynamic.
	\item Either field candidate consists of only zero bytes.
	\item One field candidate is static, the other dynamic and contains the static candidate's value in at least one message.
\end{itemize}


\section{Evaluation}\label{sec:evaluation}
We evaluate the quality of our message type identification approach by examining binary protocol traces of DNS, NBNS, NTP, SMB, and DHCP%
\footnote{
	Dynamic Host Configuration Protocol (RFC 2131), 
	Domain Name System (RFC 1035), 
	NetBIOS Name Service (RFC 1002), 
	Network Time Protocol (RFC 958), and
	Server Message Block (Microsoft) 
}
and analyzing the precision $P$ and recall $R$ of identified message clusters, which are defined as:

\newcommand{\tp}{\ensuremath{\text{TP}}}
\newcommand{\fp}{\ensuremath{\text{FP}}}
\newcommand{\tn}{\ensuremath{\text{TN}}}
\newcommand{\fn}{\ensuremath{\text{FN}}}

$$P = \frac{\tp}{\tp + \fp} \text{\quad and \quad} R = \frac{\tp}{\tp + \fn}$$

For multiple clusters, $\tp$, $\fp$, $\tn$, and $\fn$, constituting the so called confusion matrix, are defined through the correct and incorrect assignments of messages, as discussed by \citet{manning_introduction_2009}.
Positives (\_P) therefore are all pairs of messages that are classified into the same cluster, while such a pair is true (TP) if both messages are of the same type.
Negatives (\_N) and false (F\_) assignments are defined accordingly.
The amount of positives and negatives for $n$ clusters $c_i$ are thus given as:
$$\tp + \fp = \sum_{i} \binom{ \left| c_i \right| }{2} \qquad \text{and} \qquad \tn + \fn = \sum_{i, j} \left( \left| c_i \right| \cdot \left| c_j \right| \right),$$
where $j = \lbrace 0 \dots (n-1) \rbrace \setminus i$.
The number of true positives and false negatives (and, through the above equations, the complete confusion matrix) are given by:
$$\tp = \sum_{i} \sum_{l} \binom{\left| t_{i,l} \right|}{2} ~\text{and}~\fn = \sum_{i} \sum_{l} \frac{( \left| t_l \right| - \left| t_{i,l} \right| ) \cdot \left| t_{i,l} \right| }{2},$$
where $t_{i,l}$ are the messages of type $l$ in cluster $i$, while $t_l$ denotes the messages of type $l$ and thus $t_l = \cup_{i} \  t_{i,l}$.

\medskip
To obtain the required ground truth for testing our results against, we implemented Python modules that obtain the true field dissection of each message and compare these to the inferred message types.
As ground truth about the protocol specification, we utilize tshark\footnote{Command line interface of Wireshark, see \url{www.wireshark.org}}. 
For each message in the trace, we compare the inference results to the according protocol dissector provided by tshark.
%
As specimens, we use the binary protocols DNS, NBNS, NTP, SMB, and DHCP.
We chose these protocols as representatives of different typical binary protocols.
DHCP and SMB have varying amounts and lengths of fields.
We chose NTP because it is a protocol of fixed field lengths, where the lengths range from 1 to 8 bytes for the different fields.
DNS and NBNS contain mostly 2 byte binary fields mixed with variable length fields of ASCII-encoded characters.
The traces we analyzed are publicly available%
\footnote{
	NTP, NBNS, SMB, and DHCP filtered from \url{download.netresec.com/pcap/smia-2011/};
	\quad
	DNS filtered from \url{ictf.cs.ucsb.edu/archive/2010/dumps/ictf2010pcap.tar.gz}};
we pre-processed each raw trace by filtering for only the desired protocol, removing duplicates of the payload, and truncating them to 1\,000 messages each.

\newcommand{\dhcp}{\color{uulm-mawi}\large$\bullet$}
\newcommand{\dns}{\color{uulm-in}$\blacklozenge$}
\newcommand{\nbns}{\color{uulm-med}$\bigstar$}
\newcommand{\ntp}{\color{uulm-nawi}$\blacksquare$}
\newcommand{\smb}{\color{black}$\blacktriangle$}

\begin{figure*}[]
	
	\subcaptionbox{
		Segmenting with tshark.
		\label{fig:nemetyl-tshark}}{
		\begin{tikzpicture}[
			scale=4.5,
			every path/.style={line width = 1pt},
			every node/.style={font={\footnotesize\sffamily\scriptsize}}
		]
			\foreach \x in {4, 8}
				\foreach \y in {4, 8}
					\draw[fill=uulm-akzent!20, draw=none] (\x/10,\y/10) rectangle +(0.2,0.2);
			\foreach \x in {3, 6, 7}
				\foreach \y in {3, 6, 7}
					\draw[fill=uulm-akzent!20, draw=none] (\x/10,\y/10) rectangle +(0.1,0.1);

			\draw[->] (0.3,0.3) to (1,0.3);
			\draw[->] (0.3,0.3) to (0.3,1);
			\node at (0.65,0.18) {precision};
			\node[rotate=90] at (0.12,0.65) {recall};
			
			\foreach \x in {3,...,10}
				\pgfmathsetmacro\result{\x * 0.1}
				\draw (\x/10,0.31) -- (\x/10,0.29) node [below] {\scriptsize \pgfmathprintnumber{\result}};
			\foreach \x in {3,...,10}
				\pgfmathsetmacro\result{\x * 0.1}
				\draw (0.31,\x/10) -- (0.29,\x/10) node [left] {\scriptsize \pgfmathprintnumber{\result}};
			
			\node at (1,0.54) {\dhcp};
			\node at (1,0.99) {\nbns};
			\node at (1,0.98) {\dns};
			\node at (1,0.79) {\ntp};
			\node at (.81,0.88) {\smb};
		\end{tikzpicture}
	}
	\hfill
	\subcaptionbox{
		Fixed segments of 4 bytes.
		\label{fig:nemetyl-4bytesfixed}
	}{
		\begin{tikzpicture}[
			scale=4.5,
			every path/.style={line width = 1pt},
			every node/.style={font={\footnotesize\sffamily\scriptsize}}
		]
			\foreach \x in {4, 8}
			\foreach \y in {4, 8}
			\draw[fill=uulm-akzent!20, draw=none] (\x/10,\y/10) rectangle +(0.2,0.2);
			\foreach \x in {3, 6, 7}
			\foreach \y in {3, 6, 7}
			\draw[fill=uulm-akzent!20, draw=none] (\x/10,\y/10) rectangle +(0.1,0.1);
			
			\draw[->] (0.3,0.3) to (1,0.3);
			\draw[->] (0.3,0.3) to (0.3,1);
			\node at (0.65,0.18) {precision};
			
			\foreach \x in {3,...,10}
			\pgfmathsetmacro\result{\x * 0.1}
			\draw (\x/10,0.31) -- (\x/10,0.29) node [below] {\scriptsize \pgfmathprintnumber{\result}};
			\foreach \x in {3,...,10}
			\pgfmathsetmacro\result{\x * 0.1}
			\draw (0.31,\x/10) -- (0.29,\x/10) node [left] {\scriptsize \pgfmathprintnumber{\result}};
			
			\node at (0.69,0.96) {\dhcp};
			\node at (0.37,0.74) {\dns};
			\node[opacity=.66, inner sep=0] (nbns) at (1,0.32) {\nbns};  
			\node[above left=0 of nbns, inner sep=0] {\color{uulm-med}(at recall 0.13)};
			\node at (0.34,0.96) {\ntp};
			\node at (0.55,0.76) {\smb};
		\end{tikzpicture}
	}
	\hfill
	\subcaptionbox{
		Segments of NEMESYS. 
		\label{fig:nemetyl-nemesys}}{
		\begin{tikzpicture}[
			scale=4.5,
			every path/.style={line width = 1pt},
			every node/.style={font={\footnotesize\sffamily\scriptsize}}
		]
			\foreach \x in {4, 8}
			\foreach \y in {4, 8}
			\draw[fill=uulm-akzent!20, draw=none] (\x/10,\y/10) rectangle +(0.2,0.2);
			\foreach \x in {3, 6, 7}
			\foreach \y in {3, 6, 7}
			\draw[fill=uulm-akzent!20, draw=none] (\x/10,\y/10) rectangle +(0.1,0.1);
			
			\draw[->] (0.3,0.3) to (1,0.3);
			\draw[->] (0.3,0.3) to (0.3,1);
			\node at (0.65,0.18) {precision};
			
			\foreach \x in {3,...,10}
			\pgfmathsetmacro\result{\x * 0.1}
			\draw (\x/10,0.31) -- (\x/10,0.29) node [below] {\scriptsize \pgfmathprintnumber{\result}};
			\foreach \x in {3,...,10}
			\pgfmathsetmacro\result{\x * 0.1}
			\draw (0.31,\x/10) -- (0.29,\x/10) node [left] {\scriptsize \pgfmathprintnumber{\result}};
			
			\node at (0.99,0.56) {\dhcp};
			\node at (0.69,0.98) {\dns};
			\node at (1.00,0.98) {\nbns};
			\node at (1.00,0.95) {\ntp};
			\node at (0.85,0.74) {\smb};
		\end{tikzpicture}
	}
	\hfill
	\vline width 1.5pt
	\hfill
	\subcaptionbox{
		Clustering quality of Netzob.
		\label{fig:netzob}
      }{
		\begin{tikzpicture}[
			scale=4.5,
			every path/.style={line width = 1pt},
			every node/.style={font={\footnotesize\sffamily\scriptsize}}
		]
			\foreach \x in {4, 8}
			\foreach \y in {4, 8}
			\draw[fill=uulm-akzent!20, draw=none] (\x/10,\y/10) rectangle +(0.2,0.2);
			\foreach \x in {3, 6, 7}
			\foreach \y in {3, 6, 7}
			\draw[fill=uulm-akzent!20, draw=none] (\x/10,\y/10) rectangle +(0.1,0.1);
			
			\draw[->] (0.3,0.3) to (1,0.3);
			\draw[->] (0.3,0.3) to (0.3,1);
			\node at (0.65,0.18) {precision};
			
			\foreach \x in {3,...,10}
			\pgfmathsetmacro\result{\x * 0.1}
			\draw (\x/10,0.31) -- (\x/10,0.29) node [below] {\scriptsize \pgfmathprintnumber{\result}};
			\foreach \x in {3,...,10}
			\pgfmathsetmacro\result{\x * 0.1}
			\draw (0.31,\x/10) -- (0.29,\x/10) node [left] {\scriptsize \pgfmathprintnumber{\result}};
			
			\node at (0.97,0.38) {\dhcp};
			\node at (0.35,1.00) {\dns};
			\node at (0.97,0.69) {\nbns};
			\node at (0.99,0.69) {\ntp};
			\node at (0.42,0.95) {\smb};
		\end{tikzpicture}
	}
	
    \caption{Clustering quality results of \nemetyl using different segmenters and Netzob for comparison.\\
    	 The protocols are {\dhcp} DHCP, {\dns} DNS, {\nbns} NBNS, {\ntp} NTP, and {\smb} SMB.}
	\label{fig:nemetyl-eval}
\end{figure*}
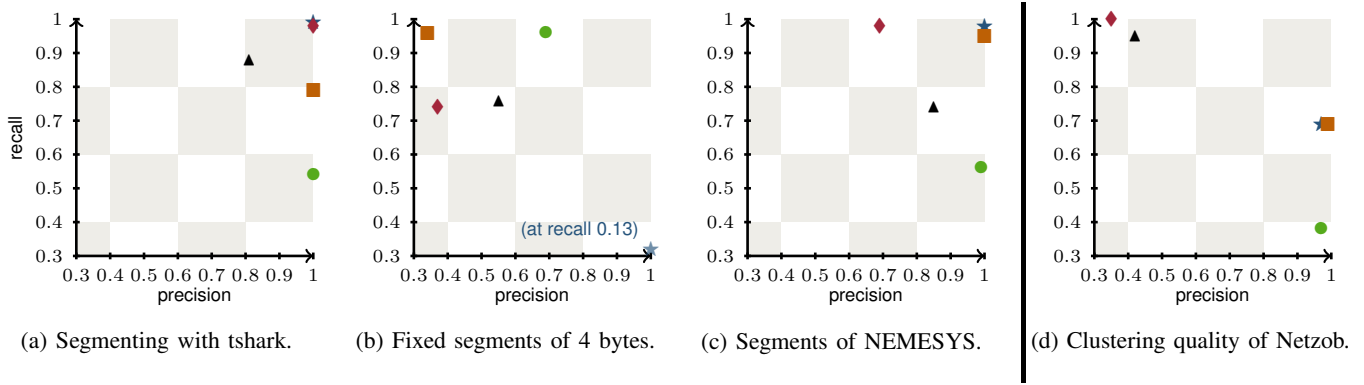

\subsection{Test Cases for Clustering Quality}
To show the validity of the assumptions that our approach is based on and the impact of the selected segmenter providing the field candidates, we 
run our implementation with different segmenters that provide all messages of a trace segmented into 
	(1) the true fields from the dissector,
	(2) equal-length chunks of 4 bytes, and
	(3) the hypothetical fields inferred by NEMESYS.
%

%

\subsection{Results}

In this section, we present the results of applying \nemetyl with the three described segmenters.
Our approach considers messages as noise if the clustering and refinement cannot determine a reliable choice for a classification.
The amount of noise for each inference is given alongside with the precision and recall.
To set our results into context, we use Netzob for inferring the same traces as we evaluate our approach with.
Netzob uses NW to byte-wise align all messages and identify message types from hierarchical clustering using the scores for this alignment.

Netzob and the segmenter NEMESYS each require to set a parameter.
The parameter $\sigma$ adjusts the Gaussian filter in NEMESYS to the expected maximum lengths of fields in the target protocol.
We used the optimal values given in \citet{kleber_nemesys:_2018}.
Netzob's parameter is the similarity threshold that needs to be reached for the hierarchical clustering to terminate and thus directly adjusts the clusters.
For Netzob, we determined the optimal similarity thresholds for each protocol in multiple test runs iterating the parameter.
The selected parameter values are given in \autoref{tab:nemetyl-eval}.

\medskip
\subsubsection{tshark Segments}\label{sec:tshark-segments}
Using true field borders as message segments, the message types of all protocols could be determined with high accuracy as can be seen in \autoref{tab:nemetyl-eval}, \textit{tshark} column, and illustrated by \autoref{fig:nemetyl-tshark},
with the exception of DHCP, which has a complex structure that does not clearly reflect the message type.
This is noticeable by the low recall value,
which is due to structurally diverse message types that are split up into multiple clusters.
SMB's relatively low precision is due to one large cluster of structurally identical messages which our splitting heuristic was not able to discriminate further.

\medskip
\subsubsection{4-bytes-fixed Segments}\label{sec:4-bytes-fixed-segments}
As to be expected, the use of an uninformed segmentation to find field candidates yields considerably worse results, as can be observed in \autoref{tab:nemetyl-eval}, \textit{4-bytes-fixed} column, and is illustrated by \autoref{fig:nemetyl-4bytesfixed}.
For protocols that determine their message type by single flag fields, in this case NBNS, this causes low recall values.
The simple field structure of DNS and NTP is obscured by the fixed segment splitting, leading to low precision values.
However, especially for the complex protocols DHCP and SMB the similarities of the messages are sufficiently recovered by this simple method to identify significant differences in message types.

\begin{table*}[]
	\centering
      \caption{Clustering quality of \nemetyl using different segmenters (columns two to four) and Netzob (column five).} \label{tab:nemetyl-eval}
	\begin{tabular}{l|rrr|rrr|rrrr||rrr}
		\emph{Segmenter} & \multicolumn{3}{c|}{\emph{tshark}} & \multicolumn{3}{c|}{\emph{4-bytes-fixed}} & \multicolumn{4}{c||}{\emph{NEMESYS}} & \multicolumn{3}{c}{Netzob method} \\
		& \multicolumn{1}{l}{precision} & \multicolumn{1}{l}{recall} & \multicolumn{1}{l|}{noise} & \multicolumn{1}{l}{precision} & \multicolumn{1}{l}{recall} & \multicolumn{1}{l|}{noise} & \multicolumn{1}{l}{precision} & \multicolumn{1}{l}{recall} & \multicolumn{1}{l}{noise} & \multicolumn{1}{l||}{$\sigma$} & \multicolumn{1}{l}{precision} & \multicolumn{1}{l}{recall} & \multicolumn{1}{l}{threshold}\\ \hline 
		DHCP \vphantom{\Large M} & 1.00 & 0.54 & 35 & 0.69 & 0.96 & 0 & 0.99 & 0.56 & 7 & 0.6 & 0.97 & 0.38 & 78 \\ 
		DNS & 1.00 & 0.98 & 18 & 0.37 & 0.74 & 17 & 0.69 & 0.98 & 7 & 0.6 & 0.35 & 1.00 & 50 \\ 
		NBNS & 1.00 & 0.99 & 21 & 1.00 & 0.13 & 21 & 1.00 & 0.98 & 13 & 0.9 & 0.97 & 0.69 & 58 \\ 
		NTP & 1.00 & 0.79 & 6 & 0.34 & 0.96 & 2 & 1.00 & 0.95 & 27 & 1.2 & 0.99 & 0.69 & 57\\ 
		SMB & 0.81 & 0.88 & 8 & 0.55 & 0.76 & 16 & 0.85 & 0.74 & 65 & 1.2 & 0.42 & 0.95 & 55\\ 
	\end{tabular}
\end{table*}


\medskip
\subsubsection{NEMESYS Segments}\label{sec:nemesys-segments}
Finally, NEMESYS infers field boundaries heuristically.
In \autoref{tab:nemetyl-eval}, \textit{NEMESYS} column, we give the quality and the sigma values used for each protocol to configure NEMESYS. 
We illustrate the results by \autoref{fig:nemetyl-nemesys}.
It can be noticed that DHCP and SMB have comparably low recall results.
This is due to the complexity of the protocol that not always reflects message types by their structure.
Like for the tshark segmenter, more than one cluster is created for one message type, although the other types are identified correctly.
The high correctness of the field boundaries determined by NEMESYS for NBNS and NTP causes the excellent quality of these protocols' inference.
NEMESYS even outperforms tshark regarding NTP recall due to one large cluster that our heuristic splitting severs falsely in the case of tshark.

\medskip
\subsubsection{Netzob Baseline}



For comparison, we apply the inference of Netzob to the same protocol traces as our approach.
The resulting clusters' precision and recall, as illustrated in \autoref{fig:netzob}, can be found in the last column of \autoref{tab:nemetyl-eval}, alongside the similarity threshold used for each protocol.

\medskip
Unsurprisingly, \nemetyl in conjunction with the ground truth segmenter tshark clearly outperforms Netzob.
In a more realistic case where the message segmentation is unknown,
\nemetyl with the NEMESYS segmenter yields a lower recall for SMB compared to Netzob and significantly outperforms Netzob for DHCP, DNS, NBNS, NTP, and SMB in terms of precision.
In summary, \nemetyl in combination with NEMESYS constitutes a relatively reliable method of high-quality to identify unknown protocol's message types that exhibit structural differences.

\subsection{Interpretation}
What sets \nemetyl apart from previous methods of message type identification is shown in the following interpretation of the evaluation results.

\medskip
(1) The high quality of applying our implementation to dissector-generated true fields (\autoref{sec:tshark-segments}) shows:
\begin{itemize}[noitemsep]  
	\item 
	Distinctive differences between 
	field types exist and can be exploited to derive similarity between messages.
	\item 
	The feature (byte value vector) and dissimilarity measure (based on Canberra) we employ works well in revealing these differences.
	\item 
	Our interpretation of mixed-dimension vectors reflects real similarities between mixed-length segments.
	\item 
	Our combination of alignment with a continuous, i.\,e., non-discrete, 
	segment-similarity matrix for calculating alignment-match scores constitutes a valid method.
\end{itemize}

\medskip
(2) Applying our implementation to field-structure-agnostic equal-length segments of messages (\autoref{sec:4-bytes-fixed-segments}) shows:
\begin{itemize}[noitemsep]  
	\item 
	Our approach works to some extent even when no knowledge about field boundaries is available.
	\item 
	Having fixed-length segments allows for a simple and efficient dissimilarity calculation providing quick results, especially for complex protocols.
	\item 
	The dissector-generated fields for the tshark and NEMESYS segmenters required us to conceive a novel method to calculate dissimilarities between mixed-length segments.
	In contrast, using fixed lengths of segments in this second stage of the evaluation allows us to calculate distances by the well-known Canberra distance within the constant vector space of four dimensions.
	Comparing these alignment results to those of the tshark and NEMESYS segmenters, which use our Canberra dissimilarity, shows that Canberra dissimilarity provides valid dissimilarity measurements and thereby backs the validity of the mixed-length dissimilarity.
\end{itemize}

\medskip
(3) Applying our implementation to NEMESYS-inferred segments of messages that denote hypothetical fields (\autoref{sec:nemesys-segments}) shows:
\begin{itemize}[noitemsep]  
	\item 
	The quality of the segmentation influences the alignment and type identification.
	\item 
	Non-trivial 
	segmentation is possible without a-priori knowledge about the protocol, in turn reinforcing the results of our previous work NEMESYS.
	\item 
	Thus, for a fully automated tool, NEMESYS and \nemetyl are a feasible combination of methods.
\end{itemize}

\subsection{Limitations}\label{sec:limitations}

A general limitation of static traffic analysis is that it is prevented by encryption of the messages.
This can only be overcome by obtaining a plain-text trace \cite{duchene_state_2018}.
For example, a Man-in-the-Middle between two genuine entities can record the decrypted messages in transit \cite{fereidooni_breaking_2017-1}.
This method requires control over the network topology.
\medskip

Depending on the protocol trace to be analyzed, in particular complex message structures lead to overspecific clusters.
Our approach is focused on determining message types on structural similarities.
Despite the cluster refinements we propose to overcome this, a protocol that defines distinct but structurally very similar message types can still lead to clusters of mixed message types.
In this case, clusters that contain messages of multiple types mislead the cluster merging heuristic to merge even more misclassified messages into one mixed cluster.
In our set of test traces this was the case for DHCP and SMB.
As we have shown, even in this case, \nemetyl is superior to previous alignment-based approaches like Netzob.

\section{Future Work}\label{sec:futurework}
Besides the three evaluated segmenters, other methods of determining atomic chunks of network messages could be devised.
However, known methods that may be suited for this task originate from natural language processing \cite{wang_semantics_2012} and require adaption to be applicable to binary protocols.

Another line of research is to confirm that our approach can be applied to other protocols without significant parameter tuning.
To this end, we plan to validate the robustness of \nemetyl with more known and unknown protocols.

\section{Conclusion}\label{sec:conclusion}
In this paper, we presented a novel method of message type identification for unknown network protocols.

\nemetyl aligns messages without relying on identical byte values to determine the similarity of field candidates or single bytes.
With this method we are able to efficiently identify message types of binary protocols.
The novelty of our approach is that we abstract from discrete byte values to feature vectors that allow for a similarity measure with a continuous value range.
Thus, we are able to discover structural patterns, which remain hidden when only exact value matches are considered.
We use message segments, not bytes, as atomic parts of a message, and apply sequence alignment to it.
We combine Hirschberg alignment with DBSCAN as cluster algorithm to improve performance over agglomerative clustering.
This results in another benefit of the approach of \nemetyl over previous approaches: It does not require to select any parameter a-priori for the analysis of an unknown protocol.
We accomplish this by proposing methods to automatically configure all employed algorithms, particularly including DBSCAN clustering.

We evaluated our approach to validate different aspects of our solution.
To have enough information about the protocol specifications of our test traces for deriving the reverse engineering quality, we used known protocols for a quantitative comparison.
The results for three different segmentation algorithms show that \nemetyl has considerable advantages in message type identification result quality over previous approaches.
Since using a similarity measure with a continuous value range for message parts and messages to analyze an unknown protocol, our approach denotes a fundamentally new method for analyzing binary protocols.

\bigskip
\section*{Acknowledgements}
The authors would like to thank David Mödinger for his assistance in clarifying the notation of this paper.

\bigskip
\printbibliography



\end{document}